\documentclass[10pt,a4paper,final]{article}
\usepackage[a4paper, total={6.2in, 9in}]{geometry}
\usepackage[utf8]{inputenc}
\usepackage{amsmath}
\usepackage{amsfonts}
\usepackage{amssymb}
\usepackage{authblk}
\usepackage{graphicx,wrapfig}
\usepackage{hyperref}
\title{{\bf Strongly Correlated Electrons in Solids}}
\author{Henri Alloul \thanks{Henri Alloul (2014), {\it Strongly Correlated Electrons in Solids}, Scholarpedia, 9(7):32067.}}
\affil{LPS - CNRS/Universite Paris Sud, Orsay, France}
\date{}

\begin{document}
\maketitle
\begin{abstract}
Most emergent properties of the materials discovered since the 1980s are related to the existence of electron-electron interactions which are large with respect to the kinetic energies and could not be thoroughly studied before. The occurrence of  metal insulator transitions, of exotic magnetic and/or superconducting properties in many new compounds have stimulated a large series of experimental and theoretical developments to grasp their physical significance. We present here a simple introduction to the elementary aspects of the physics of electron-electron interactions, which could be a starting point for typical undergraduate students. 
\end{abstract}

\section{Introduction}
\label{Introduction}
The study of the electronic properties of solids, done within an independent electron approximation since World War II, has been essential for the understanding of the occurrence of semi-conductors. This understanding was at the origin of the information technologies which expanded rapidly after the war. But during that period, a myriad of new materials with increasing complexity have been discovered as well. These materials were found to display unexpected novel electronic properties. Many such properties are not explained by the independent electron approximations, require new conceptual developments, and will certainly lead in the future to specific promising applications. Most of these emergent properties are linked with magnetic responses due to the strong electron-electron interactions in these complex new materials.

We will briefly discuss how these electronic interactions yield original states of electronic matter. A variety of experimental and theoretical techniques have been developed which permit a detailed investigation of their unexpected properties.

This article will be organised as follows. Electronic properties of solids were, in the first half of the twentieth century, considered mostly in the frame of an independent electron approximation with spin degeneracy. The resulting electronic band structure of metals which will be briefly recalled in section \ref{Band_struc_basics} is such that each electronic level could be doubly occupied. In such an approach one expects metals or insulators with no significant magnetic properties.

In order to explain that some solids display magnetic properties one must reassess the underlying approximations that led to the band theory, and especially the averaging approach to the Coulomb interactions between electrons. In section \ref{Atomic_mag_orign} we shall show that one has to take into account the strong local coulomb repulsion on atomic orbitals, which permits magnetic atomic states and magnetic insulators in the solid state.

We shall then specifically mention in section \ref{Superconductivity} the superconducting state which is an original correlated electronic state which occurs in most metals at low temperature. This is a macroscopic quantum electronic state which results from an indirect electron-electron attractive interaction induced by the interplay of electronic and atomic vibrations in classical metals in which the electron states do not interact at high temperatures.

We shall then consider in section \ref{Mott_Super} how electronic correlations yield materials with properties which are in an intermediate regime between independent, delocalised electrons and local states. Those intermediate electronic states are at the basis of the correlated electron physics. They often display exotic superconducting states with unexpectedly high transition temperatures, can undergo charge ordering or metal insulator transitions as well as exotic magnetic states considered as spin liquids. Such original states, which are far from being fully understood at this time, will be introduced in dedicated Scholarpedia articles. 

\section{The basics of the electronic band structure of solids}
\label{Band_struc_basics}

\begin{figure}
\begin{center}
\includegraphics[height=7cm,width=8.5cm]{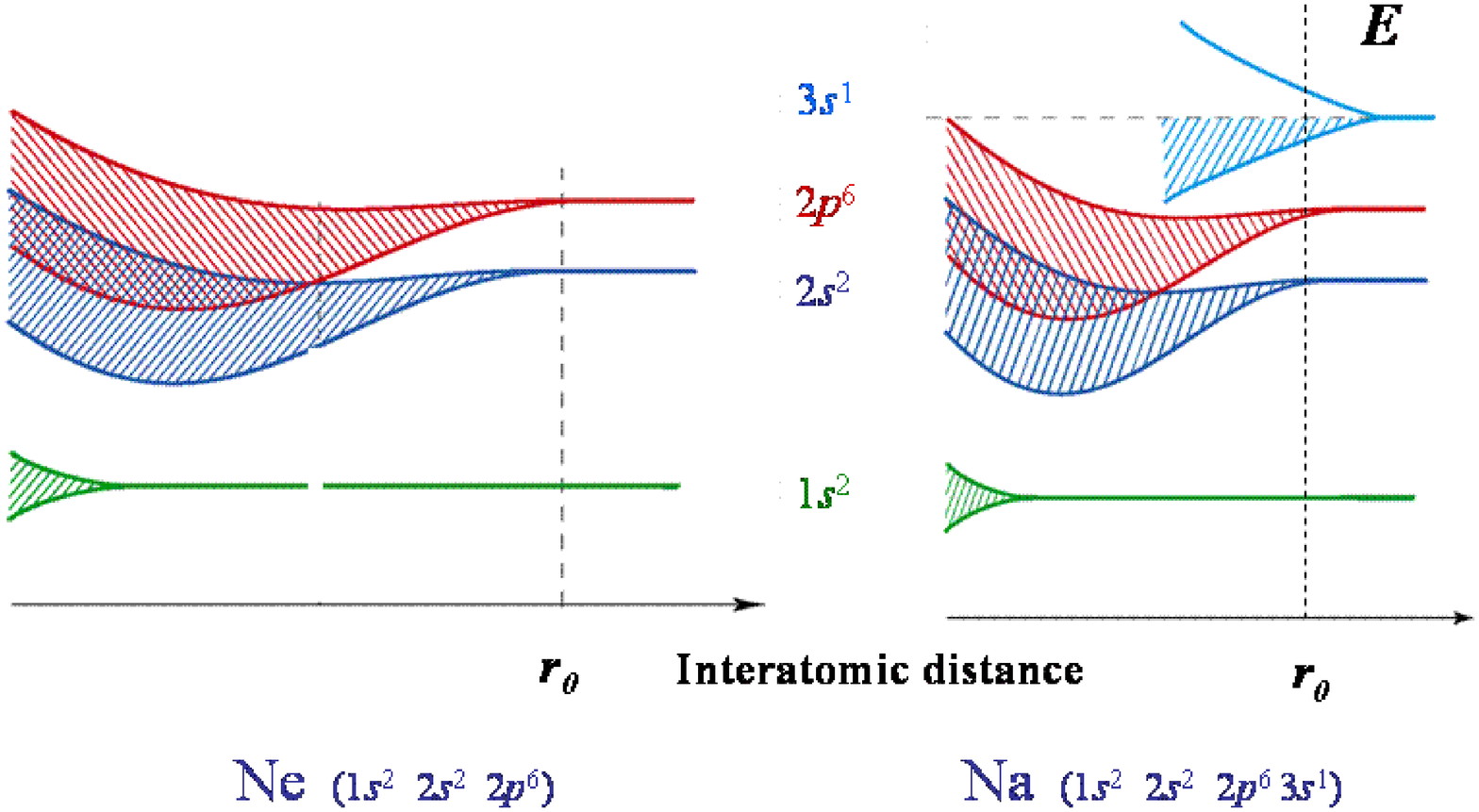}
\caption{\label{fig:Fig1}Band structures of $Ne$ and $Na$. Here the calculated energy bands are plotted versus a fictitious distance between atomic orbitals and $r_0$ represents the equilibrium distance at ambient temperature. Ne only displays filled or empty bands and is an insulator, while Na has its higher energy band only filled with one electron per atom and is a metal.} 
\end{center}
\end{figure}

Isolated atoms display discrete, narrow electronic levels. In the solid state, the electron can delocalise between sites due to the overlap of the electronic orbitals of neighboring atoms. The transfer integrals $t$ between orbitals of neighboring atoms lead to a broadening of the atomic levels into electronic bands which characterize the actual band structure of a given material. The width of these energy bands is typically determined by $zt$ where $z$ is the number of neighboring atoms surrounding a given site. In such an independent electron approach the available electrons in the material fill the energy levels in increasing energy order. This yields insulators when filled and empty bands are separated by finite gaps, and metals if there are partially filled energy bands up to an energy level which defines the Fermi energy, as shown in Fig.\ref{fig:Fig1}. In such an approach, solids with an odd number of electrons per unit cell are expected to be metals as they should necessarily display partially filled bands in which delocalisation of electrons can be done at moderate energy cost. One distinguishes then among the insulators the cases where the energy gap is small compared to the thermal energy $k_BT$. In that case electrons can be excited thermally at temperature $\sim T$ into the first empty band (conduction band) and leave empty holes in the last occupied band (the valence band), this being the case of semiconductors. Among those, graphene has been highlighted recently, as in that case the gap vanishes and the conduction and valence band touch each other at a single energy point, the Dirac point which corresponds in that case to the Fermi energy.

In those cases the band theory for the electronic states applies rather well and explains most of the electronic properties of these metals, insulators, semiconductors or Dirac point metals. In all those cases the independent electrons approach yields a weak paramagnetism as all these descriptions do not lift the spin degeneracy of the electronic states. This Band theory describes these materials well as the k space construction lifts the site degeneracy of the atomic state by building Bloch states which have different energies and well defined properties under translation. 

\section{The origin of Atomic magnetism and Mott insulators}
\label{Atomic_mag_orign}

\begin{figure}
\begin{center}
\includegraphics[height=7cm,width=8.5cm]{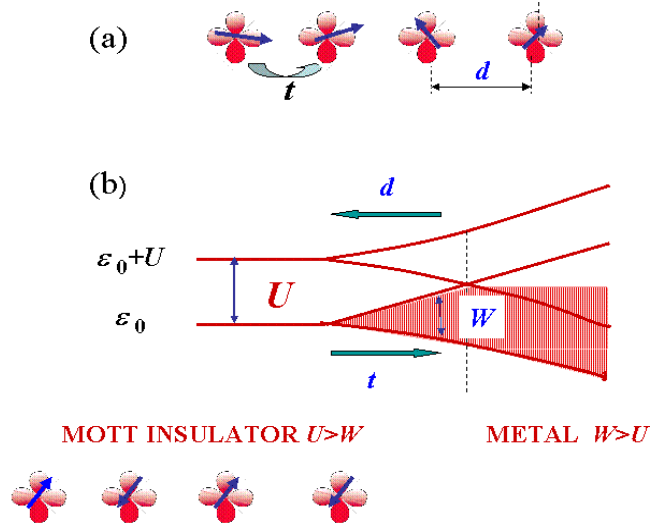}
\caption{\label{fig:Fig2}The Mott–Hubbard model. (a) The atomic orbitals on atomic sites at a distance d, with the transfer integrals t between neighboring sites . (b) The levels for the Hubbard model as a function of t or d. On the left: Isolated atom with energy levels $\epsilon_0$ and $\epsilon_0+U$. Center: Mott–Hubbard insulator obtained for a small hopping integral $W<U$. Right: Metallic situation corresponding to $W>U$. This very simple approach goes by the name of the Hubbard model.} 
\end{center}
\end{figure}

If $t$ is small, then one expects very narrow bands and localized electronic states, as the case $t=0$ corresponds to strictly isolated atomic states. In that case electronic interactions can no longer be treated as an average as done in band theories and do give rise to local moment and magnetism, as we shall see hereafter. 

\subsection{Mott insulators}
Let us begin by considering the case of an isolated atom (on the left in Fig\ref{fig:Fig2}). In this context, in band theory, it is assumed that the energy brought to the system by an extra electron would be $\epsilon_0$, and that a second electron on the same atom would also bring $\epsilon_0$, so that the total energy would be $2\epsilon_0$ for a doubly negatively charged ion. But this is obviously not very realistic, owing to Coulomb repulsion. Apart from its 'orbital' energy $\epsilon_0$, the second electron will also be subject to the Coulomb repulsion of the first electron, and its energy will thus be higher than $\epsilon_0$ by an amount usually denoted by U, which represents the Coulomb repulsion between the first and second electrons added to an initially neutral atom. The total energy of the doubly negative ion is thus $\epsilon_0 + U$. Note that $U$ can vary considerably depending on the atom (from about 1 eV to more than 10 eV).

If we now consider this ion in a crystal, the hopping integrals between nearest neighbors will broaden the discrete atomic levels into bands of width $W=zt$. To begin, we consider the limiting case of small hopping integral compared with $U$. We find ourselves in a situation corresponding to the middle of Fig.\ref{fig:Fig2}. There are two allowed energy bands called the upper and lower Hubbard bands, separated by a band gap. This gives the impression that we have a typical insulator (or semiconductor). But this is not in fact correct. There is one additional one-electron state per atom, so that, in a solid comprising $N_n$ atoms, the lower band of the middle column can contain up to $N_n$ electrons, rather than up to $2N_n$ electrons, as is the case in the context of the independent electron band theory. In particular, if there is now one electron per atom (or more generally an odd number of electrons per primitive cell), the lower band will be completely filled and the upper band completely empty. We will thus have an insulator with an odd number of electrons per primitive cell, as a consequence of the interactions $U$ between electrons. The very existence of such an insulator (usually called a Mott-Hubbard insulator in recognition of the two British scientists who first studied them in the 1960s) is thus a consequence of the Coulomb interaction between electrons. As we shall see later, important examples of Mott-Hubbard insulators are undoped cuprates in which the $Cu^{2+}$ ions are in a $3d^9$ state. 

\subsection{Magnetism of Mott Insulators}
While usual band insulators should be nonmagnetic (or more precisely, slightly diamagnetic), very different expectations occur for a Mott-Hubbard insulator. If we begin by considering the limiting case of very small hopping integrals, we end up with isolated atoms. The electron in the level $\epsilon_0$ can then have spin up or spin down, behaving like an isolated spin $1/2$. In the solid, these spins taken together will give rise to Curie paramagnetism with a spin susceptibility $\sim 1/T$, that is, a paramagnetic insulator susceptibility that contradicts band theory. If one takes into account the finite value of the hopping integral $t$, it can be shown that, at low enough temperatures, the spins on neighboring atomic sites will like to arrange themselves in opposite directions, that is, antiferromagnetic coupling dominates.

The main conclusion which can be taken here is that going beyond the possibilities offered by band theory (paramagnetic metals and diamagnetic insulators), the presence of Coulomb interactions between electrons, if they are strong enough, can give rise to an insulating state with a variety of magnetic properties, such as Curie paramagnetism, antiferromagnetism (but also ferromagnetism), and so on as will be shown later on.

The Hubbard model, which replaces the true Coulomb potential $V(r)\sim 1/r$ by a repulsion which only acts if the two electrons are located on the same atom, is clearly a drastic simplification of the actual physical situation. However, it is rather naturally justified in the context of the theory of magnetic phenomena. Experiments show that there are not only magnetic insulators of spin $1/2$ (in fact these are in the minority), but that in most cases the spin per atom is much higher. This is due to the fact that, in almost all cases, the atomic orbitals involved are not $s$ levels (hence non-degenerate), but $d$- or $f$-type (hence five- or seven-fold degenerate).

In such a situation one has to take in more detail the local repulsive Coulomb interaction between electrons on the orbitals of such poly-electronic atoms. Though the Coulomb interaction is purely electrostatic, it differentiates the energy levels of the atomic orbitals, depending of their orbital symmetry and disfavors then double occupancy of some of them. These electronic interactions when combined with the Pauli principle are responsible for the local moment magnetism of isolated atoms. In this situation, the angular momentum of each atom in the Mott-Hubbard insulating state is determined by the electronic filling of the atomic levels through specific rules named Hund's rules.

The interactions between those local moments in ordered solids are responsible for the various long range ordered magnetic states (ferromagnetic or antiferromagnetic) or their absence thereof in the case where ordering is prohibited by geometric frustration effects, as will be illustrated later on. 

\section{Superconductivity and electron-phonon interaction}
\label{Superconductivity}
So far we have seen that the original properties of electronic matter are mostly governed by the magnitude of the Coulomb repulsion between electrons which is essential in the magnetic properties and in promoting localized electronic states rather than extended states. But, although we mentioned it at many places already, we did not consider so far one of the most important correlated electronic states which has been studied at length during most of the last century, that is superconductivity. This electronic state of matter is by no way an independent electron case, as the basic feature of this state is an electronic organisation which emphasizes pairs of electrons, the Cooper pairs. This has been highlighted in classical metals by the development of the Bardeen Cooper Schrieffer (BCS) theory which states that in the presence of an attractive interaction between electrons, no matter how weak it may be, the electron gas becomes unstable. 

\begin{figure}
\begin{center}
\includegraphics[height=7cm,width=8.5cm]{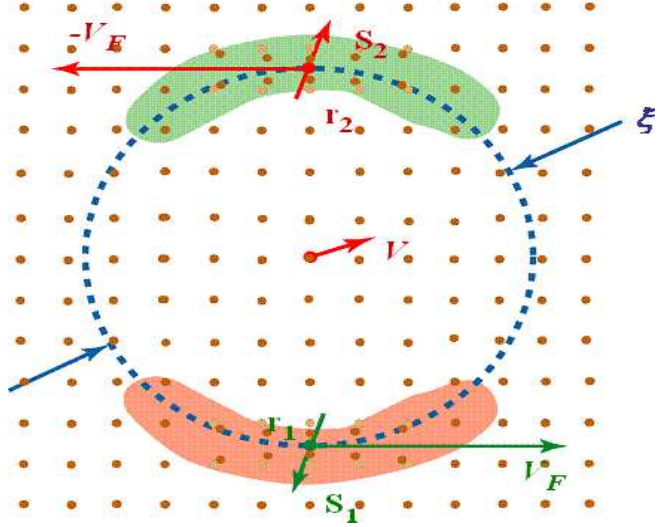}
\caption{\label{fig:Fig3}Spatial representation of a Cooper pair. The lattice distortions shown by pink and green shadings have been produced by the electrons at r2 and r1 respectively and trap the electrons with opposite spins in a singlet state.} 
\end{center}
\end{figure}
The beauty of this unexpected physical situation is that the lower energy condensed electronic state is a quantum state of electronic matter in which the correlations between electrons extend on macroscopic distances.The mystery which prevented the actual understanding of superconductivity during the first half of the 20th century concerned the actual possibility of such an attractive interaction between electrons. This has only been understood when it had been noticed that the electrons do attract the ions of the atomic background, and that their displacements (the phonons) being slow due to the large ionic masses provide a memory effect which mediates an attractive interaction between electrons. If that electron-phonon interaction dominates the electronic Coulomb repulsion, then the net attractive interaction favors the pairing of electrons which is qualitatively depicted in Fig.\ref{fig:Fig3}. The pairing of electrons results in the many body electronic states which is the basis of the electronic properties of the superconductors. One of the main unexpected behaviors which could be explained by the BCS theory is the existence of a gap between the electronic superconducting ground and excited states. The occurrence of such a gap has been initially ascertained by NMR experiments. 

\section{From Mott Insulators to Metallic Magnetism and Superconductivity}
\label{Mott_Super}
We have examined so far two completely different limiting descriptions of electronic states in a solid. In the band structure approach we have described the case of electrons considered as independent, their interactions being restricted to an averaged potential. The delocalisation of these electrons between the atomic sites driven by the transfer integrals may yield metallic states. In contrast we have considered the specific situation for which electrons localized on ionic states lead to local atomic magnetic moments. Those arise when the Pauli principle and on site inter-electronic Coulomb repulsion are taken into account properly. We have assumed implicitly that these electrons do not delocalise when the transfer integrals between electrons on neighboring ions are small enough in such solids. This then corresponds to an insulating magnetic state quite different from the band insulating states considered so far in the independent electron band approach.

The actual situation in real materials does indeed sometimes correspond to these limiting cases, but a wide variety of solids correspond to intermediate situations, like that of ferromagnetic metals such as $Fe$ or $Ni$. But the correlated electron physics is now rich with examples of such intermediate cases which are quite important both for the fundamental questions raised and for the applications of the novel physical effects which come into play. 

\subsection{From Mott insulators to metal insulator transitions}

In a Mott-Hubbard insulator, if we increase $t$ (or if we consider compounds with lower values of $U$), for a certain critical value of $t/U$, the upper and lower Hubbard bands begin to overlap (see Fig.\ref{fig:Fig2} right), causing the band gap to disappear and leading to a metallic state. Such an increase in $t$ can be produced by bringing the atoms closer together. This was first achieved in the case of doped semiconductors by increasing the donor concentration, e.g., by increasing the concentration of phosphorus in silicon. This causes the hydrogen-like orbitals of $P$ to move much closer together and increases the hopping integrals, while remaining in a configuration corresponding to one electron per donor atom. A simpler way to achieve this situation directly without changing the number of electrons in a material is to apply an external pressure. This increases the hopping integrals t by bringing the atoms closer together, provided that the material is compressible. In the metallic state thereby induced, one then observes magnetic and thermodynamic properties which require taking into account the existence of the strong coulomb repulsion $U$. As for the Mott-Hubbard insulator, let us point out that it looks at first glance like a band insulator, the only difference being that here each Hubbard band contains only Nn states rather than $2N_n$ states in the case of the band theory of section \ref{Band_struc_basics}. 

\subsection{Doping a Mott insulator: the cuprate problem}

\begin{figure}
\begin{center}
\includegraphics[height=7cm,width=8.5cm]{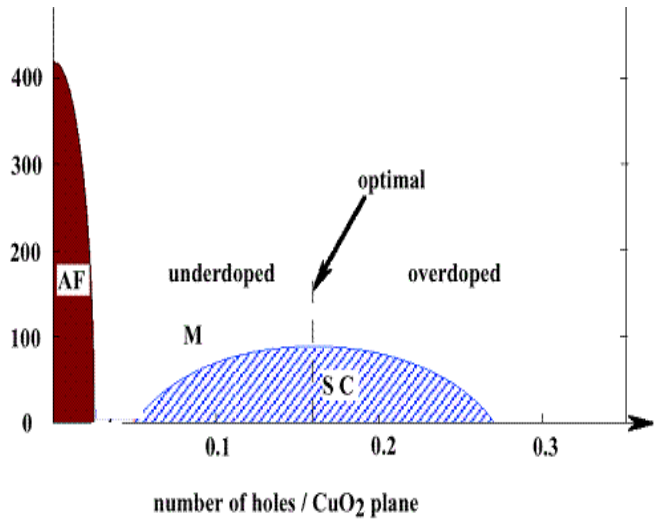}
\caption{\label{fig:Fig4}Elementary phase diagram of the cuprates obtained by hole doping the $AF$ Mott insulating state. The $AF$ state is rapidly destroyed by hole doping beyond 0.05 and opens the way to a metallic state which becomes superconducting at low $T$. One observes a SC $T_c$ dome shape as a function of the hole doping. The highest $T_c$ occurs for optimal doping and one speaks of underdoped and overdoped states.} 
\end{center}
\end{figure}

Chemical treatment may be envisaged to change the number of electrons in a Mott insulator. For example, it can be doped with holes, reducing the number of electrons in the lower Hubbard band to a number $N_{e}$ smaller than $N_n$. This is exemplified by the case of cuprates such as $YBa_2Cu_3O_6$ or $La_2CuO_4$ which are antiferromagnetic Mott insulators.In the latter, the $Cu$ are in a $3d^9$ state with spin $1/2$, which order antiferromagnetically below 340K. By chemical exchange of a fraction $x$ of $La^{3+}$ by $Sr^{2+}$ one can typically reduce the number of $Cu$ electrons to become $N_e = (1-x)N_n$. This reduction of the number of electrons in the lower Hubbard band suggests that the doped Mott-Hubbard insulator is expected to be a metal. Experimental investigations carried out on the cuprates, and also on certain other classes of doped Mott insulators, have shown that doping gradually reduces the N\`eel temperature of the antiferromagnetic state. This $AF$ state is completely suppressed for a low level of doping, of the order of $x\approx 0.05$, as can be seen in the phase diagram for $La_{2-x}Sr_xCuO_4$ displayed in Fig.\ref{fig:Fig4}.

\subsection{Superconductivity in correlated electronic systems}

The importance of the cuprates in the physics of correlated systems has resulted from the discovery that when the $AF$ is suppressed by hole doping, the doped metallic state which results has a $SC$ ground state and displays strange metallic and magnetic properties. The most surprising feature has been the fact that the superconductivity discovered in these materials has the highest critical temperatures $T_c$ found so far in any superconducting material, and exceeds any $T_c$ which could be expected within the BCS approach known to apply in classical metallic states. This has immediately led to the idea that $SC$ in the cuprates has an exotic origin linked with electron-electron interactions rather than the classical electron-phonon driven superconductivity which prevails in classical metals. An important observation in the cuprates has been the fact that the phase diagram with increasing hole doping displays a dome shaped $SC$ regime, that is $SC$ disappears for dopings beyond about 0.3.

While the cuprates are certainly exotic superconductors, let us state that many other materials have been shown to display situations where magnetism and $SC$ are proximate to each-other in phase diagrams. In pnictides those are sometimes spanned by doping as in the cuprates, but in other families of compounds the phase diagrams are spanned by pressure control of the overlap integrals as for organic, heavy fermions or $Cs_3C_{60}$ compounds. The author shall present many examples of such families in the Scholarpedia article NMR in strongly correlated materials[\ref{see}. 

\section{Experimental techniques}
\label{Exp_tech}
Such original states have been revealed initially by experimental techniques which were quite adapted at the time of the discovery of the cuprates to studies of their electronic properties. Among those, Nuclear Magnetic Resonance (NMR) is a technique which is quite essential as it permits local measurements in the materials. This gives precious information which goes beyond the first indications given by the macroscopic magnetic measurements as they permit one to differentiate the properties of the materials which can be attributed to specific phases or sites in the structure. Also, as usual for magnetic materials, inelastic and elastic Neutron scattering techniques reveal the occurrence of magnetic responses and of their k-dependence.

Significant effort has been invested to improve the quality of single crystals which are essential for the studies of the transport properties in these exotic metals and $SC$. Static or pulsed high magnetic field sufficiently large to suppress the superconducting state have been achieved, though this not yet possible for samples with high optimal $T_c$.

Other new specific techniques for studies of surfaces of 2D compounds have been developed during the last decades. The Angular Resolved Photoemission Spectroscopy (ARPES) uses X-rays generated by synchrotrons to perform k-space resolved spectroscopy of the occupied electron states. This permits determination of the band structures of these correlated electron materials. Deviations with respect to simple band calculations permit determination of the incidence and strength of the electronic correlations. Also Scanning Tunneling Microscopy experiments reveal spatial inhomogeneities of the gaps and of the electronic structures at surfaces in these materials. Some experimental groups have developed Fourier transformations at a level of refinement which allowed them to reproduce some of the ARPES spectral information. The existence of charge density wave transitions is also detected by Resonant Inelastic X-ray Scattering (RIXS) or Resonant Elastic X-ray Scattering (REXS).

Many of these novel techniques have been improved by recent technical developments, but their input on the physics of correlated electron systems are still far from being fully understood at this time and will be introduced in dedicated Scholarpedia articles[\ref{see}]. 

\section*{Refereces}
\begin{enumerate}
\item Alloul H., ``Introduction to the physics of Electrons in Solids" Graduate texts in Physics, Springer–Verlag (Heidelberg) (2011), ISBN 978-3-642-13564-4 DOI:10.1007/978-3-642-13565-1

\item Ashcroft Neil W., Mermin N. David, Saunders College (1976), ISBN 0030493463, 9780030493461

\item Kittel C., ``Introduction to Solid State Physics", 8th Edition, Wiley (2005), ISBN 047141526X, 9780471415268

\item Mott N., ``Metal Insulator transitions" Taylor \& Francis (1974), ISBN 0850660793, 9780850660791

\item Tinkham M., ``Introduction to Superconductivity", Dover Publications (1996), ISBN 0486134725, 9780486134727 
\end{enumerate}

\section*{See Also}
\label{see}
\begin{enumerate}
\item ``NMR in strongly correlated materials'', H. Alloul, Scholarpedia, 10(1):30632 (2015)

\item ``Bardeen-Cooper-Schrieffer theory", Leon Cooper and Dimitri Feldman (2009), Scholarpedia, 4(1):6439. 
\end{enumerate}

\end{document}